\documentclass[runningheads]{llncs}
\usepackage[utf8]{inputenc}
\usepackage[T1]{fontenc}
\usepackage{graphicx}
\usepackage{float}
\usepackage{svg}
\svgsetup{inkscapelatex=false}
\usepackage{amsmath} 
\usepackage{microtype}
\usepackage{xurl}
\newcommand{\maha}{\ensuremath{\text{MAHA}}}
\newcommand{\amaha}{\ensuremath{\text{anti-MAHA}}}
\usepackage{color}
\usepackage{xcolor}
\usepackage{tabularx}
\usepackage{booktabs}
\usepackage{array}
\usepackage{multirow} 
\usepackage{ulem}

\usepackage{caption}
\usepackage{hyperref} 

\makeatletter
\def\section{\@startsection{section}{1}{\z@}{-8pt plus -1pt minus -1pt}{6pt plus 1pt minus 1pt}{\normalfont\Large\bfseries}}
\makeatother

\renewcommand{\arraystretch}{0.9}
\makeatletter
\def\subsection{\@startsection{subsection}{2}{\z@}{-10pt plus -2pt minus -2pt}{6pt plus 1pt minus 1pt}{\normalfont\large\bfseries}}
\makeatother

\begin{document}

\title{Hidden in Plain Sight: Finding MAHA on Reddit}
\author{Sabit Ahmed\inst{1} \and
Subigya Nepal\inst{1} \and
Henry Kautz\inst{1}}

\authorrunning{Ahmed et al.}

\institute{University of Virginia, Charlottesville, VA 22903, USA
\email{\{bcw3zj,sknepal,rmw7my\}@virginia.edu}}

\maketitle              
\begin{abstract}
Make America Healthy Again (MAHA) is a national health movement that encompasses a striking mix of beliefs, from broadly accepted concerns about good diet and exercise to 
controversial takes on organic and genetically modified food, childhood vaccination, science, and institutions.
Various influencers and promoters of the MAHA movement on social media are scattered throughout the online space. Investigating the structure, discourse, and contagion of MAHA beliefs requires large-scale fine-grained digital footprints.
Constructing structured data covering different MAHA themes from vast unstructured social media data is challenging.
We introduce a Reddit dataset that spans six years (2020–2025), comprising 18.5M posts from 4M users. Containing the natural and thematic context of 12 MAHA-aligned beliefs, this dataset offers researchers from various domains the opportunity to study the dynamics of the MAHA movement, its structural and functional components, and the linguistic and behavioral patterns of its proponents.

\keywords{Social media \and Large-scale data \and Health discourse \and Make America Healthy Again (MAHA) \and Large language models (LLMs)}
\end{abstract}

\section{Introduction}
Make America Healthy Again (MAHA) is a national health movement that encompasses many distinct themes, ranging from uncontroversial ones, such as the importance of exercise, to controversial ones, such as increasing the consumption of red meat, to dangerous ones, such as the claim that childhood vaccinations cause autism.  
Social networking sites are one of the key components for the growth of the MAHA community.  Reddit is the largest social media platform where posts are organized thematically, i.e., by subreddits.  Therefore, it would seem natural to study the rise and spread of MAHA on Reddit by analyzing the appropriate subreddits.  
However, three challenges arise: the largest MAHA-related subreddit (r/NoNewNormal) was banned, and many similar communities were quarantined \cite{brodkin2021reddit}; Reddit is highly dynamic with 1,500--1,700 new subreddits created daily \cite{NewSubredditsMonth}; and the platform generates hundreds of millions of posts per year \cite{reddit_2024_10k}.

Therefore, we cannot restrict our attention to a small set of subreddits to find MAHA on Reddit. MAHA discourse is spread through thousands of different subreddits, so we must search Reddit as a whole. Secondly, we must develop a highly efficient pipeline to sift through terabytes of data and accurately identify the MAHA community that is ``hidden in plain sight'' on Reddit.
With a multi-stage keyword search and machine-learning pipeline, we narrowed billions of Reddit submissions and comments down to millions with the strongest MAHA signals across 12 themes. Using a tree-based, few-shot Large Language Model (LLM) pipeline, we demonstrate an analytics pipeline to characterize each user's belief themes and stances.

\section{Data Collection Pipeline}

\subsection{Reddit Data Archive Download}
\begin{figure}[t]
\includegraphics[width=\textwidth]{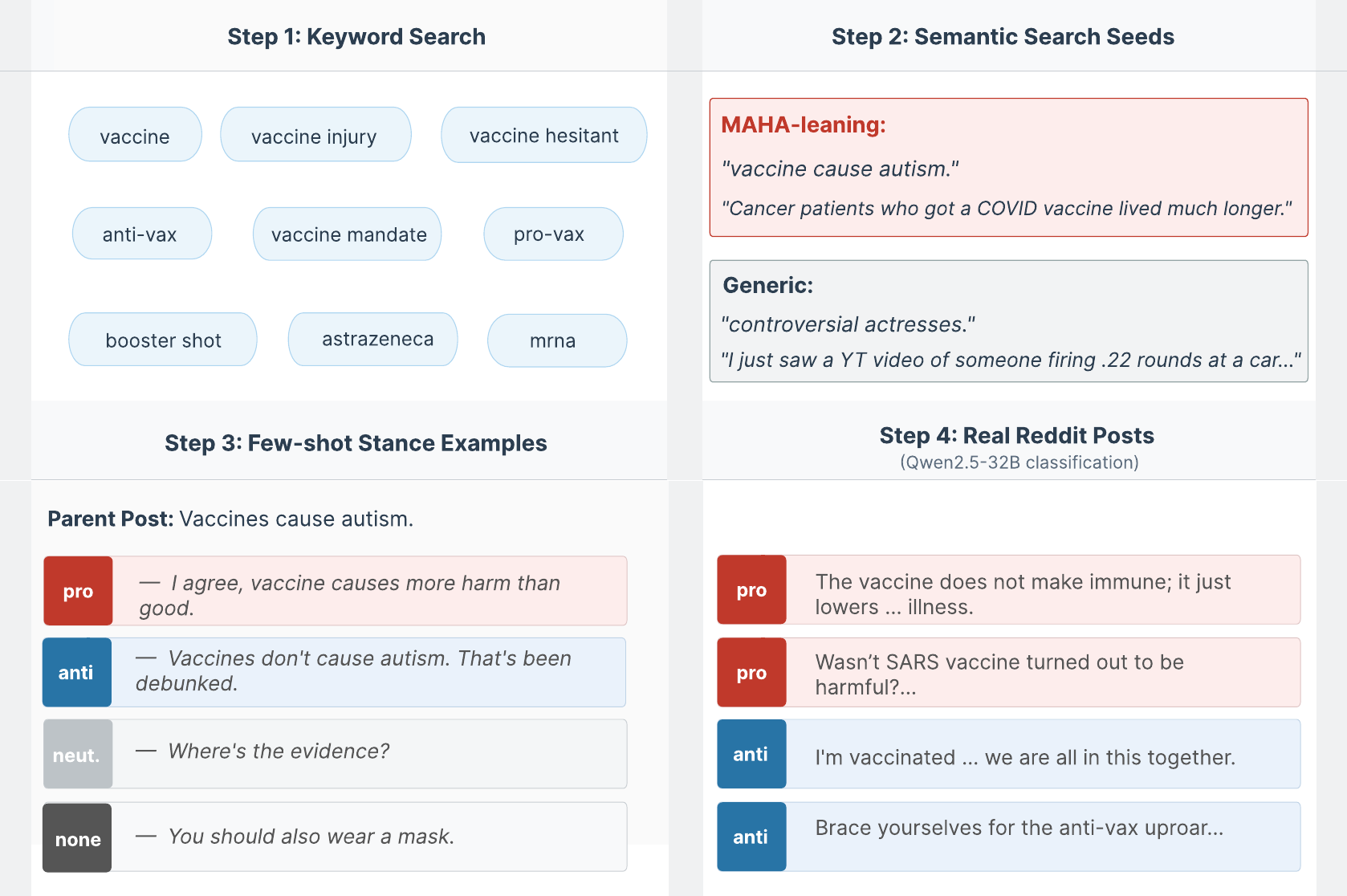}
\centering
\caption{Data collection and stance-labeling pipeline with worked examples for the Vaccine theme across four stages.} \label{fig:pipeline}
\end{figure}

Historically, the Reddit platform offered access to rich natural-language text exchanged across subreddit communities through the Pushshift API -- an endpoint widely used for large-scale behavioral research. After Reddit discontinued access to Pushshift, we obtained historical archives via Arctic Shift, a third-party system that mirrors Pushshift's data. We collected all submissions and comments and retained six years (January 2020 -- December 2025), a window that begins with the COVID-19 pandemic and captures the key period of MAHA-adjacent discourse. We use \textit{post} to refer to both submissions and comments. An overview of the pipeline is shown in Fig.~\ref{fig:pipeline}.

\subsection{Defining Themes of MAHA communities}
After careful assessment of recent studies and the report published by the White House \cite{alba2026maha,xue2026makebeefgreatagain,WhiteHouse2025_MAHA}, we consider 12 MAHA-related themes that span lifestyle (e.g., exercise, food),
controversial health topics (e.g., vaccines, fluoride), and broader ideological
domains (e.g., anti-science, climate skepticism).
\begin{itemize}
\item \textbf{Exercise:} Promotion of exercise and physical fitness.
\item \textbf{Food:} Processed food criticism and promotion of whole foods.
\item \textbf{GMO:} Skepticism toward genetically modified organisms and crops.
\item \textbf{Fluoride:} Opposition to public water fluoridation.
\item \textbf{Vaccine:} Criticism of vaccines (e.g., vaccine injury).
\item \textbf{Mask:} Opposition to mask mandates in the context of COVID-19.
\item \textbf{Pharma:} Criticism of industry and evidence-based medical practice.
\item \textbf{Science:} Distrust of mainstream scientific consensus and institutions.
\item \textbf{EMF:} Unproven effects of electromagnetic field (EMF) radiation.
\item \textbf{Environment:} Concerns about pollutants (e.g., microplastics, toxic chemicals, pesticides).
\item \textbf{Climate:} Skepticism of mainstream climate science.
\item \textbf{Screen:} Concerns about the health effects of phone and computer use.
\end{itemize}
Each theme can contain pro-MAHA, anti-MAHA, or neutral opinion stances. The distinction between stances is demonstrated in Fig. \ref{fig:pipeline} with example posts.

\subsection{2-Stage Keyword Search and Cleaning}
We performed a 2-stage keyword search with a list of keywords and regular expressions. For the first stage, we combined all 12 sets of keywords into a single regular expression to search through $\sim$20 billion submissions and comments. 
In the second stage, we applied the search to each individual keyword set and obtained 12 distinct sets of thematic data. 
Following this, we removed any non-English texts and posts from bots and admins. We kept only the posts that had more than 3 words. This resulted in 234M submissions and 1.3B comments in total. 

\subsection{Classifying MAHA-only Discussions}
Keyword-matched data still contained many posts irrelevant to MAHA -- e.g., ``shot'' captures both vaccine and gun discussions. Therefore, we trained a logistic regression classifier on MAHA relevance. 
We randomly sampled 200 submissions and 200 comments per theme per month over six years (2020–2025), yielding ~28,800 samples per theme. Using 100 theme-specific seed queries, we ran semantic search on each thematic sample, sorted those by cosine similarity, and kept high-scoring ones ($\geq 0.9$) as candidate positives. We manually reviewed the top $\sim$3,000 candidates per theme, yielding $\sim$14,000 MAHA-positive samples overall. For negatives, we applied a low similarity threshold ($<0.2$) along with the MAHA-negative seed queries, obtaining $\sim$68,000 MAHA-negative samples; a manual inspection confirmed this similarity-based labeling was effective and captured a wide variety of generic discussions.
We then applied hard negative mining to replace easy negatives with the 20,000 most confusing examples around the decision boundary.
We evaluated two algorithms on two types of features: logistic regression with Sentence-BERT embeddings (LR + SBERT) and a support vector machine with TF-IDF-weighted unigrams, bigrams and trigrams (SVM + N-grams), using 10-fold cross-validation. LR + SBERT achieved the highest recall and F$_1$ and was selected as the final classifier (precision $=0.982$, recall $=0.974$, F$_1=0.978$). The high F$_1$ reflects the relative separability of MAHA-relevant posts from irrelevant posts in the semantic embedding space; a baseline classifier without hard negative mining achieved F$_1=0.970$, indicating that hard negative mining contributed a small but consistent improvement rather than driving the headline performance. The selected model was then applied to all posts matched with keywords to retain only MAHA-relevant content for downstream analysis. Details of the classifier performance are reported in Table~\ref{tab:maha_classifier}.
\begin{table}[t]
    \centering
    \caption{10-fold cross-validation results for MAHA vs.\ generic classification. 
    F1\textsuperscript{*} denotes the positive class (MAHA).}
    \label{tab:maha_classifier}
    \begin{tabular}{llccccc}
    \toprule
    \textbf{Negatives} & \textbf{Model} & \textbf{AUC} & \textbf{Prec.} & \textbf{Rec.} & \textbf{F1\textsuperscript{*}} & \textbf{F1\textsubscript{Mac}} \\
    \midrule
    \multirow{2}{*}{Baseline}
      & LR + SBERT      & \textbf{.999} & .959 & \textbf{.982} & .970 & \textbf{.982} \\
      & SVM + $n$-grams & .997          & .973 & .949          & .961 & .976 \\
    \midrule
    \multirow{2}{*}{Hard-neg}
      & LR + SBERT      & .997          & \textbf{.982} & .974 & \textbf{.978} & .981 \\
      & SVM + $n$-grams & .992          & .981          & .935 & .957          & .963 \\
    \bottomrule
    \end{tabular}
\end{table}
\subsection{Identifying Themes and Stances from Posts}
After obtaining the most relevant MAHA posts across all 12 themes, we applied two post-processing steps: removing submissions with fewer than 10 comments and removing all comments without parent submissions. This resulted in 1.2M submissions and 45M comments mapped to one or more of the 12 themes.
Reddit discussions are organized as a tree-based thread, where people discuss or criticize a theme-specific issue through submissions, comments, or replies. We identified four challenges in classifying the stances from a post:
First, a post that passed through the  initial logistic-regression-based MAHA filter may not substantively discuss any theme, containing only spurious signals.
Second, a post can take a clear stance, no stance, or contain a mix of opinions, questions, or information on a specific theme, either as a parent (i.e., submission, comment) or as a child (i.e., comment, reply).
Third, a child comment or reply can agree or disagree with its parent post or may not take a clear side.
Fourth, a post may discuss multiple themes and stances, requiring a predictive model to assign an independent stance to each relevant theme.

To obtain relevant themes and stances from a post, we used a tree-based few-shot in-context learning. We provided the LLM with a few-shot example to predict the themes and stances of each post. 
We first applied a relevance gate, using an LLM to determine whether a post was actually related to any MAHA (Make America Healthy Again) theme, filtering out false positives that passed the initial logistic regression classifier. Posts that passed this filter were then classified for theme and associated stance.
We defined three stances a post could take toward a theme — pro, anti, and neutral — allowing the model to assign different stances to different themes within the same discussion. 

\begin{table}[ht]
    \centering
    \caption{LLM theme and stance classification performance on the evaluation set.
    Relevance F$_1$ measures binary theme presence;
    Theme F$_1$ is macro-F$_1$ across the 12 themes;
    and Stance F$_1$ is macro-F$_1$ across pro, anti, and neutral stances.
    $^\dagger$Selected as the final production model.}
    \label{tab:stance_classifier}
    \small
    \begin{tabular}{lccc}
    \toprule
    \textbf{Model} & \textbf{Relev.\ F1} & \textbf{Theme F1} & \textbf{Stance F1} \\
    \midrule
    Qwen2.5-7B          & 0.67  & 0.48  & 0.42  \\
    Qwen2.5-14B        & \textbf{0.78}  & \textbf{0.60}  & 0.54  \\
    Qwen2.5-32B$^\dagger$ & 0.76  & \textbf{0.60}  & \textbf{0.61}  \\
    \bottomrule
    \end{tabular}
\end{table}

We evaluated multi-theme-stance classification performance using the Qwen2.5 model family. Using stratified sampling across the 12 themes, we selected 500 submission trees, then sampled 500 comments from the pooled set of comments across those trees.
One researcher annotated all 1000 samples and a second researcher independently annotated a stratified subsample of 100 to measure the interrater agreement.
Agreement (Cohen's $\kappa$) was substantial for both theme ($\kappa$ = 0.86) and stance ($\kappa$ = 0.75) classification. The best performing model achieved an F1 of 0.60 for theme classification and 0.61 for stance classification and was selected for large-scale theme and stance labeling.
The resulting data comprises $\sim$478K submissions and $\sim$18.1M comments. Table \ref{tab:dataset_summary} shows the statistics on the curated data. The data will be available for download upon request from the authors' website \href{https://www.cs.virginia.edu/~bcw3zj/projects/MAHA/data/}{(link)}.
The data analyzed in this study are publicly available on Reddit and were accessed in accordance with the platform’s terms and conditions. We report analyses exclusively at the aggregate community level; therefore, no privacy concerns apply.

\begin{table}[ht]
\centering
\caption{Dataset Summary}
\begin{tabular}{lr}
\hline
\textbf{Metric} & \textbf{Count} \\
\hline
Unique submissions        & 478,626 \\
Unique comments           & 18,120,340 \\
Unique users              & 3,964,483 \\
Unique subreddits         & 18,096 \\
\hline
\end{tabular}
\label{tab:dataset_summary}
\end{table}
\subsection{Identifying User Stances and Distinctive Words}
To obtain a fine-grained signal for downstream analyses, retained users who had at least three posts in every year from 2020 to 2025. This resulted in a total of 33,384 users.
As an illustrative application of the constructed dataset, we identified user stances and extracted distinctive words.
We characterized each user by their fraction of posts in each stance across all themes, using a weighted opinion scoring, adjusted from Wu et al. \cite{wu_twitter_2025}:

\begin{figure}[ht]
\centering
\includegraphics[width=0.95\columnwidth]{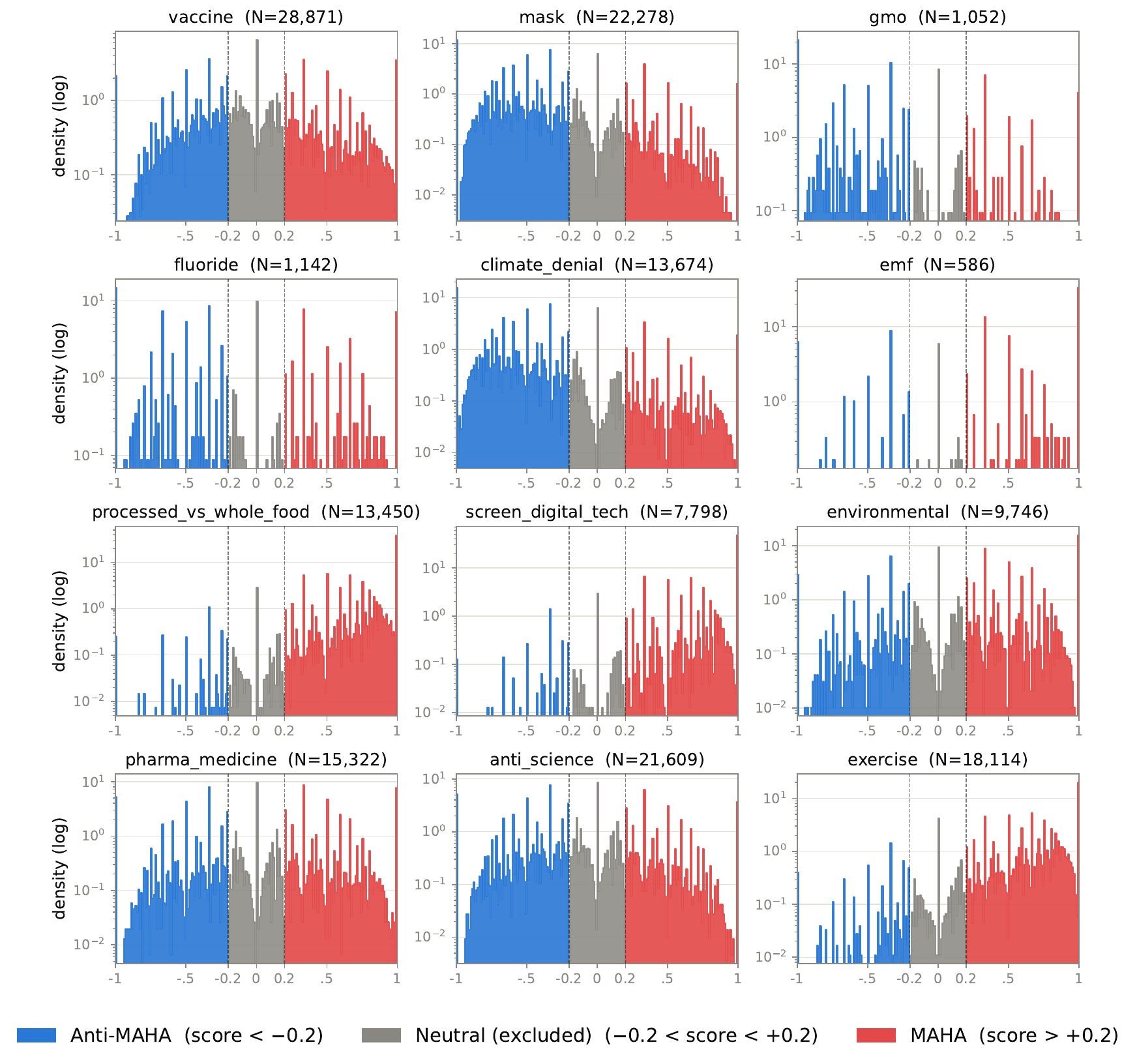}
\caption{User stance distribution on 12 themes.} \label{fig:fig_opinion_score}
\end{figure}

\begin{equation}
  S(u,t) =
    \frac{w_{s}\,(n_{\text{pro}}^{s} - n_{\text{anti}}^{s})
        + w_{c}\,(n_{\text{pro}}^{c} - n_{\text{anti}}^{c})}
         {w_{s}\,n_{s} + w_{c}\,n_{c}}
  \label{eq:op_score}
\end{equation}

Where $n_{\text{pro}}^{s}$, $n_{\text{anti}}^{s}$, $n_{\text{neutral}}^{s}$ denote stance counts from submissions, and $n_{\text{pro}}^{c}$, $n_{\text{anti}}^{c}$,
$n_{\text{neutral}}^{c}$ from comments,
$n_{s}$ and $n_{c}$ are the total
submission and comment counts,  
and $w_{s}$ and $w_{c}$ are the weights for a user $u$ and theme $t$,
\begin{equation}
  w_{s} = \frac{n_{c}}{n_{s} + n_{c}}, \qquad
  w_{c} = \frac{n_{s}}{n_{s} + n_{c}}
  \label{eq:op_weights}
\end{equation}
Scores range from $-1$ (anti-MAHA) to $+1$ (MAHA). 
After careful inspection on the distribution of opinion scores as shown in Figure \ref{fig:fig_opinion_score},
we label users as \maha{} if $S \geq 0.2$ or \amaha{} if $S \leq -0.2$; users with $|S| < 0.2$ are excluded from subsequent analyses, as their net stance signal is too weak to attribute a clear position. 
After obtaining the \maha{} and \amaha{} user groups, we statistically matched the unique users with four covariates: posts per year, number of active themes, mean word count, and mean percentage of function words. 
We then calculated a differential TF-IDF score for each word as the difference in the mean TF-IDF between the \maha{} and \amaha{} groups, and extracted the top five unique and most distinctive words per group for all 12 themes. Table~\ref{tab:top_words} shows that \maha{} users are more likely to criticize institutions (e.g., ``monsanto'', ``big pharma'', ``cdc'', etc.), power dynamics and economic structures (e.g., ``government'', ``companies'', ``money'') and exhibit negative sentiment against vaccines, masks and conventional medical practices (e.g., ``forced'', ``control''). By contrast, \amaha{} users emphasize evidence and scientific consensus (e.g., ``evidence'', ``scientific'', ``research''), are more likely to trust medical advice and treatment (``vaccinated'', ``effective'', ``advice'', ``treatment''), and use critical words towards non-scientific claims (``conspiracy'', ``anti-science'', ``stupid''). 
This pattern reflects not only the ideological divide between the two groups, but also the internal diversity of topics within the MAHA community itself, such as discussions linking vaccine skepticism with 5G -- a unique conspiracy narrative \cite{flaherty2022conspiracy} stemmed from two otherwise distinct threads of MAHA. 
\begin{table*}[ht]
\vspace{-6pt}
\centering
\scriptsize
\caption{Top 5 unique, distinctive words.}
\renewcommand{\arraystretch}{1.5}
\setlength{\tabcolsep}{4pt}

\makebox[\textwidth][c]{%
\begin{tabularx}{1.08\textwidth}{
    >{\raggedright\arraybackslash}p{1.35cm}
    >{\raggedright\arraybackslash}X
    >{\raggedright\arraybackslash}X
}
\hline
\textbf{Theme} &
\textbf{MAHA-distinctive words} &
\textbf{Anti-MAHA distinctive words} \\
\hline

Vaccine &
\textcolor{red}{vaccine, trump, pharma, mrna, government} &
\textcolor{blue}{vaccinated, doses, risk, available, effective} \\
\midrule

Mask &
\textcolor{red}{mask mandates, lockdowns, government, evidence, forced} &
\textcolor{blue}{vaccinated, sick, n95, risk, indoors} \\
\midrule

GMO &
\textcolor{red}{monsanto, roundup, anti, vaccine, mrna} &
\textcolor{blue}{better, non-gmo, grown meat, selective breeding, organic} \\
\midrule

Fluoride &
\textcolor{red}{iq, gland, conspiracy, water supply, rfk} &
\textcolor{blue}{teeth, dentist, use, toothpaste, dental} \\
\midrule

Climate &
\textcolor{red}{co$_2$, global warming, science, temperature, man} &
\textcolor{blue}{action, vote, fossil, republicans, need} \\
\midrule

EMF &
\textcolor{red}{vaccine, gates, covid, chip, control} &
\textcolor{blue}{radio, radiation, frequency, stupid, ionizing} \\
\midrule

Food &
\textcolor{red}{sugar, fiber, weight, keto, carbs} &
\textcolor{blue}{evidence, processed, plant, science, vegan} \\
\midrule

Screen Use &
\textcolor{red}{algorithms, social media, dopamine, internet, content} &
\textcolor{blue}{screen, kid, tv, vr, games} \\
\midrule

Environment &
\textcolor{red}{food, lead, water, eat, chemicals} &
\textcolor{blue}{energy, carbon, emissions, nuclear, tax} \\
\midrule

Pharma &
\textcolor{red}{big pharma, vaccine, pfizer, money, companies} &
\textcolor{blue}{medication, adhd, advice, need, treatment} \\
\midrule

Science &
\textcolor{red}{government, covid, virus, media, cdc} &
\textcolor{blue}{evidence, scientific, conspiracy, anti-science, research} \\
\midrule

Exercise &
\textcolor{red}{workout, training, started, good, home} &
\textcolor{blue}{weight, calories, exercise, weight loss, diet} \\
\hline

\end{tabularx}%
}

\label{tab:top_words}
\end{table*}
\section{Applications}
The released dataset enables a broad range of analyses on multi-theme belief communities. Per-theme stance labels and per-user aggregate scores support \textbf{sentiment analysis} of how MAHA-aligned and mainstream users frame health topics, both within and across themes. The coexistence of mainstream and contested themes in a single dataset enables study of \textbf{polarization and narratives} --- in particular, how supporters and detractors construct competing accounts of the same topic through polarized lenses, and how narratives diffuse between mainstream and fringe themes. Reddit's reply structure, preserved in our dataset, supports analysis of \textbf{network structure and community}, including interaction graphs between user groups, sub-community formation around specific themes, and cross-theme co-engagement patterns. The six-year window (2020--2025), which spans the COVID-19 pandemic and the rise of MAHA as a political movement, supports \textbf{temporal analysis} of belief adoption, theme-to-theme transitions, and shifts in stance composition over time. Finally, the large volume of stance-labeled text enables fine-grained \textbf{linguistic analysis}, including psycholinguistic profiling, framing differences, and characteristic vocabulary across user groups and themes. 
Detailed analyses and relevant applications of the constructed dataset is thoroughly discussed in a companion paper which is currently under review~\cite{ahmed2026online_movement}.

\section{Related Work}
The MAHA movement originated as a small community of vaccine skeptics, critics of pesticides and toxic chemicals, and proponents of healthy and organic food under the banner of ``health freedom'' \cite{stolberg2026maha}. 
However, it grew to a powerful national movement with the appointment of Robert F. Kennedy Jr. as the United States Secretary of Health and Human Services and took control of many federal health policies. Numerous studies examined online discourse to understand public sentiments and narrative shifts towards vaccines and anti-science attitudes. Gyawali et al. \cite{gyawali_shifting_2025} investigated how the perception of English-speaking X (former Twitter) users evolves using millions of posts on vaccine discourse. 
Wu et al. \cite{wu_twitter_2025} found that Twitter communities are associated with users' opinion shifts towards the COVID-19 vaccine in Japan. 
In a Twitter-based analysis, Rao et al. \cite{rao_political_2021} examined the correlation between political partisanship and anti-science attitudes.
Paino et al. \cite{paino_medical_2024} examined factors that affect digital health literacy.
Dalege and Van Der Does \cite{dalege_using_2022} proposed a cognitive network model to investigate interrelated beliefs on childhood vaccination and genetically modified (GM) food.
Fariello and Jemielniak \cite{fariello_changing_2025} analyzed 16 years of Reddit discussions that span 11.5 billion posts to examine the language and sentiment surrounding climate change.
Recently, Charles Alba \cite{alba2026maha} leveraged pre-trained language models (PLMs) to understand sentiments and topics commonly discussed in X's (former Twitter) MAHA community.

\section{Conclusion and Future Work}
We introduced a large-scale Reddit dataset to study the Make America Healthy Again (MAHA) movement, covering six years (2020--2025) and 12 MAHA-aligned belief themes. Using a three-stage pipeline that combines keyword search, a relevance classifier, and a few-shot LLM stance classifier, we labeled 478K submissions and 18.1M comments from 4M users by per-theme stance and aggregated them into per-user stance scores, yielding MAHA, anti-MAHA, and mainstream user groups.
Keyword-based filtering, though two-stage and validated, may miss MAHA-aligned engagement expressed implicitly --- for example, brief agreement comments (``agreed'', ``same here'') that inherit stance from their parent posts but lack MAHA vocabulary. More context-aware filtering could recover such cases.
By preserving the natural and thematic context of MAHA discourse at scale, the dataset opens avenues for studying the structural, temporal, linguistic, and behavioral dimensions of contemporary health movements online.

\bibliographystyle{splncs04}
\bibliography{bibtex}

\end{document}